\documentclass[12pt,letterpaper]{article}
\usepackage[includeheadfoot,
            marginratio={1:1,2:3},
            width=412pt,
            height=688pt,]{geometry}
\usepackage[usenames,dvipsnames,svgnames,table]{xcolor}
\usepackage{amsmath}
\usepackage{amsfonts}
\usepackage{amssymb}
\usepackage{graphicx}
\usepackage{color}

\usepackage{empheq}

\newcommand{\nc}{\newcommand}
\nc{\beq}{\begin{equation}}
\nc{\eeq}{\end{equation}}
\nc{\bea}{\begin{eqnarray}}
\nc{\eea}{\end{eqnarray}}
\def\ov{\overline}
\def\cO{{\cal O}}
\def\IP{\mathbb{P}}
\newcommand{\drawsquare}[2]{\hbox{%
\rule{#2pt}{#1pt}\hskip-#2pt
\rule{#1pt}{#2pt}\hskip-#1pt
\rule[#1pt]{#1pt}{#2pt}}\rule[#1pt]{#2pt}{#2pt}\hskip-#2pt
\rule{#2pt}{#1pt}}

\newdimen\csize\csize=1.5ex
\def\young#1{\tiny\vcenter{\hbox{\vrule\vtop{\hrule
  \offinterlineskip\halign{&\vbox
  {\hbox to\csize {\strut\hss##\hss\vrule}\hrule}\cr#1 \crcr}}}}}

\newcommand{\Ysymm}{\raisebox{-.5pt}{\drawsquare{6.5}{0.4}}\hskip-0.4pt%
        \raisebox{-.5pt}{\drawsquare{6.5}{0.4}}}
\newcommand{\Yasymm}{\raisebox{-3.5pt}{\drawsquare{6.5}{0.4}}\hskip-6.9pt%
        \raisebox{3pt}{\drawsquare{6.5}{0.4}}}

\newcommand{\eq}[1]{\begin{equation}
                     \begin{split} #1 \end{split}
                     \end{equation}}

\begin{document}

\vspace*{-1.5cm}
\begin{flushright}
  {\small
  MPP-2012-87\\
  }
\end{flushright}

\vspace{1.5cm}
\begin{center}
  {\LARGE
A Note on  Poly-Instanton Effects in Type IIB\\[0.2cm]
Orientifolds on Calabi-Yau Threefolds}

\end{center}

\vspace{0.75cm}
\begin{center}
  Ralph Blumenhagen$^1$, Xin Gao$^{1,2}$, Thorsten Rahn$^1$ and
Pramod Shukla$^1$
\end{center}

\vspace{0.1cm}
\begin{center}
\emph{
$^{1}$ Max-Planck-Institut f\"ur Physik (Werner-Heisenberg-Institut), \\
   F\"ohringer Ring 6,  80805 M\"unchen, Germany\\
$^{2}$ State Key Laboratory of Theoretical Physics, Institute of Theoretical
Physics,\\ Chinese Academy of Sciences, P.O.Box 2735, Beijing 100190, China } \\[0.1cm]

\vspace{0.2cm}

 \vspace{0.5cm}
\end{center}

\vspace{1cm}


\begin{abstract}
The zero mode structure for the generation
of poly-instanton corrections for Euclidian $D3$-branes
wrapping complex surfaces  in  Type IIB orientifolds with $O7$- and $O3$-planes
is analyzed.
Working examples of such surfaces and explicit
embeddings into compact Calabi-Yau threefolds are presented,
with special emphasis on geometries
capable of realizing   the LARGE volume scenario.
\end{abstract}

\clearpage



\section{Introduction}
\label{sec:intro}

For connecting  string theory to our four-dimensional world the understanding
of moduli stabilization and its consequences for the effective
four-dimensional theory is of utmost importance, both for particle physics and
cosmology. During the last decade we have seen many advances in
identifying  mechanisms how moduli stabilization can be achieved, like
non-trivial  background fluxes \cite{Gukov:1999ya,Taylor:1999ii,Giddings:2001yu}, perturbative corrections
\cite{Becker:2002nn,Berg:2007wt} and instanton effects \cite{Witten:1996bn}. A couple of
principle mechanisms, which often invoke a combination of these effects, have
been identified and extensively discussed in all its consequences in the
literature.  This includes the KKLT \cite{Kachru:2003aw} and the LARGE
volume scenario \cite{Balasubramanian:2005zx}, which
are best understood for Type IIB orientifolds with $O7$ and
$O3$-planes. The rules of string model building for
such backgrounds have been worked out in
\cite{Blumenhagen:2008zz,Collinucci:2008sq,Cicoli:2011it,Cicoli:2011qg}
and a couple of important steps  toward honest fully fledged
realistic string models have been done.
Nevertheless,
we think it is fair to say that from a top-down perspective no honest  string
compactification has been proven to give rise precisely to these
effects. Clearly, this has to do with the large number of moduli and with the
intricate relations and constraints  for the various objects and structures
present in genuine string models. Here we have in mind,
constraints from tadpole cancellation, Freed-Witten anomalies \cite{Freed:1999vc} or correlations between the presence of fluxes/D-branes and
the instanton zero mode
structure \cite{Blumenhagen:2007sm}.

Although  the concept of inflation has been proposed quite some time ago  as a
solution to certain cosmological issues \cite{Guth:1980zm,Linde:1981mu},
the embedding of inflationary scenarios into a semi-realistic model in string
theory has been convincing  only after all the moduli could be stabilized. Such
an inflationary model has been initiated in \cite{Kachru:2003sx} in which,
following the idea of \cite{Dvali:1998pa}, an
open string modulus appearing as a brane separation was argued to be an
inflaton candidate. There has been a large amount of work dedicated
to build sophisticated models of open string inflation (see the reviews
\cite{Quevedo:2002xw,McAllister:2007bg,Baumann:2009ni} and references therein).
However, in particular in the framework of the LARGE volume
scenario, closed string moduli inflation
has also been seriously considered
\cite{Conlon:2005jm,Conlon:2008cj,Cicoli:2008gp,Cicoli:2011zz}.
In this respect, moduli are of interest which at leading order
still have a flat potential and only by a, in the overall  theory, subleading
effect receive their dominant contribution. These can be either  subleading
perturbative or instanton effects.

Such a subleading instanton contribution is given by so-called
poly-instanton effects, which can be briefly described
as instanton corrections to instanton actions.
These were introduced and studied
in the framework of Type I string compactifications in
\cite{Blumenhagen:2008ji}.
The analogous poly-instanton effects will also appear in the
aforementioned Type IIB
orientifolds with $O7$ and $O3$ planes.
Thus, these corrections seem to be quite promising from the point of
view of constructing
(semi-)realistic models for attempts to address several open issues in string
cosmology as well as in string phenomenology.

Recently, utilizing these ideas of poly-instanton effects,
 moduli stabilization and inflationary aspects have been studied in
a series of papers
\cite{Blumenhagen:2008kq,Cicoli:2011yy,Cicoli:2011ct,Cicoli:2012cy,Cicoli:2012tz}. However
the analysis had to be  carried out from  a rather heuristic point of view, as
a clear understanding of the string theoretic conditions  for the generation
of these effects was lacking. It is the aim of this note to clarify the zero
mode conditions for an Euclidian D3-brane instanton, wrapping a divisor of the
threefold, to  generate such a poly-instanton effect. We also provide concrete
examples of such divisors and present a couple of Type IIB orientifolds on
toric Calabi-Yau threefolds containing them. In this respect,  we are heading
for examples which also contain shrinkable del Pezzo surfaces so that the
LARGE volume scenario can in principle be realized.
We find that in the simplest class of Type IIB orientifold models, in
which poly-instanton corrections are guaranteed to be present, the K\"ahler
potential takes a very peculiar form which is different from those of
the fibrations  used in \cite{Cicoli:2011it,Cicoli:2011yy,Cicoli:2011ct}.
However, we would like to emphasize that the goal of our investigation is
to formulate sufficient conditions for the generation
of poly-instanton effects.
After invoking  additional effects, like gauge or
closed string fluxes, it might be possible to soak up the
additional fermionic zero modes \cite{Bergshoeff:2005yp,Blumenhagen:2007bn}.
Clearly, in such cases a more detailed
analysis of the precise intanton actions is necessary.

The paper is organized as follows: In section {\bf 2}, after recalling the structure of poly-instanton corrections to the superpotential  and some basic notions of Type IIB orientifolds, we investigate the zero mode structure of the respective $E3$ instantons wrapping complex surface in Calabi-Yau threefolds.  This leads to surfaces of a certain topology, whose embedding into concrete Calabi-Yau threefolds is studied in section {\bf 3}. After specifying the threefolds, we also identify admissible orientifold projections for the generation of poly-instanton corrections. Along with E3-instanton contributions, we also consider  the possibility of non-perturbative contributions to the superpotential coming from  gaugino condensation. Finally, the section {\bf 4} presents the conclusions.

In this paper, we rather focus on the  model building and mathematical issues and postpone the interesting analysis of moduli stabilization and cosmological applications to up-coming work \cite{BGRS}.

\section{Poly-instanton corrections}

The notion of poly-instantons was introduced in \cite{Blumenhagen:2008ji} (see \cite{{GarciaEtxebarria:2007zv,Dorey:2002ik,Petersson:2010qu}} also for earlier related work) and means the correction of an Euclidian  D-brane instanton action by
other  D-brane instantons.
The configuration of interest in the following is that
we have two instantons $a$ and $b$. The zero mode
structure of instanton $a$ is such  that it generates
a correction to the holomorphic superpotential of the form
\eq{   W= A_a\, \exp(-S_a) \, .
}
Here $S_a$ denotes the classical D-brane
instanton action and the prefactor $A_a$ a
moduli dependent one-loop determinant, which can be understood
as the exponential of the holomorphic part of the one-loop correction to the
classical instanton action. This is completely analogous
to the classical holomorphic gauge kinetic function and its one-loop
correction on a fictitious space-time filling D-brane.
Due to a non-renormalization theorem, such a gauge kinetic function
can be corrected at one-loop order and by instantons, so that
one expects the same also for the classical instanton action.
Thus, if the second instanton $b$ has the right zero mode
structure to generate such a correction, one gets
\eq{
       W&=\exp\left( -S_a + S_{a}^{\rm 1-loop}+A_b\, e^{-S_b}\right)\\[0.1cm]
     &= A_a\, \exp\left( -S_a\right) + A_a\, A_b\, \exp\left(
       -S_a-S_b\right)+\ldots\; .
}
The contribution of the instanton $b$ is clearly exponentially
suppressed relative to the contribution of the instanton $a$.
However, as we will review in a moment, the two zero mode
structures are different so that the instanton actions
$S_a$ and $S_b$ will generally depend on different
moduli. Therefore, it can happen that the leading order
occurrence of the moduli governing $S_b$ is through this
poly-instanton effect, which then has to be taken into
account for moduli stabilization.

\subsubsection*{Type IIB orientifolds}

In \cite{Blumenhagen:2008ji} the zero mode analysis was carried out for pure
$\Omega$ orientifolds, i.e. compactications of the Type I
superstring on Calabi-Yau manifolds.
In this paper,  we are interested in the case, where one compactifies the
Type IIB superstring on a Calabi-Yau threefold ${\cal M}$  and performs
an orientifold quotient   $\Omega \sigma (-1)^{F_L}$. Here,
$\sigma$ is a holomorphic involution acting on the K\"ahler form
$J$ and the holomorphic $(3,0)$-form $\Omega_{3,0}$ as
\eq{
          \sigma(J)=J\, \qquad \sigma(\Omega_{3,0})=-\Omega_{3,0}\, .
}
This leads to $O7$-and $O3$-planes as fixed point loci.
Recall that it is for this class of compactification
where moduli stabilization is understood best.
Generically, the complex structure moduli and the dilaton get
fixed by turning on a non-trivial closed string three-form flux
$G_3=F_3+\tau H_3$. At tree-level this implies a no-scale
structure leaving the K\"ahler moduli as flat directions.
These can be stabilized by Euclidian D3-brane instantons
wrapping four-cycles $E\subset {\cal M}$.

The presence of $O7$-planes wrapping a divisor $O7$
give rise to a tadpole for the R-R $C_8$-form
which has to be canceled by stacks of $D7$-branes wrapping divisors $D_a$ and
their orientifold images $D'_a$ so that
\eq{
               \sum_a N_a\, (D_a+D'_a)  =8\, O7\, .
}

There is an important  subtlety which has to do
with the Freed-Witten anomaly, appearing if the divisor $D$
wrapped by a $D7$-brane is non-spin. The
quantization condition for the gauge flux on the  $D7$-brane
reads
\eq{
      c_1(L) -i^* B +\frac{1}{2} c_1(K_D)\in H^2(D,\mathbb Z)\, .
}
where $i^*$ denotes the pull-back of  forms from the
Calabi-Yau threefold onto the divisor $D$.
For a non-spin divisor $c_1(K_D)$ is not even so that one is forced
to introduce a half-integer $B$-field or gauge flux
$c_1(L)={1\over 2\pi} {\cal F}$ with ${\cal F}=F+ i^* B$.
(see \cite{Blumenhagen:2008zz} for more
details on building Type IIB orientifolds).
For all the examples to be discussed later, we cancel the
Freed-Witten anomaly by ${\cal F}=0$ and appropriate
half-integer $B$-field backgrounds.

For a brane wrapping a 4-cycle $D$ which is invariant under
the orientifold projection $\Omega \sigma (-1)^{F_L}$
one can get orthogonal and symplectic  gauge groups.
Turning on a gauge flux with $c_1(L)\in H^{11}_-(D)$ the
fluxed brane is still invariant under the orientifold
projection. This is in contrast to a flux
$c_1(L)\in H^{11}_+(D)$ which breaks the gauge symmetry
to a unitary group.

The $O7$-planes and the $D7$-branes also induce a
$D3$-brane tadpole, which generally
reads
\eq{
N_{D3}+\frac{N_{\text{flux}}}{2}+N_{\text{gauge}}=\frac{N_{O3}}{4}+\frac{\chi(D_{O7})}{12}+\sum_a N_a\, \frac{\chi_o(D_{a})+\chi_o(D'_{a})}{48}
}
with $N_{\text{flux}}=\frac{1}{(2\pi)^4 \alpha^{'2}}\int H_3\wedge F_3 $, $N_{\text{gauge}}=-\sum_{a} \frac{1}{8\pi^2} \int_{D_a} \text{tr}{\cal F}_a^2$
and $\chi_o$ denoting the Euler characteristic of a
smoothened divisor (as motivated by duality to
F-theory \cite{Collinucci:2008pf}).
In this paper we are not so much concerned with the
physics on the $D7$-branes and for the concrete orientifolds
we will always cancel the $D7$-brane tadpole by simply
placing eight $D7$-branes on top of the $O7$-plane.
Clearly, for concrete model building this simple assumption
has to be relaxed.
Now, the $D3$ tadpole condition simplifies to
\eq{
\label{d3tadsimple}
      N_{D3} + \frac{N_{\text{flux}}}{2}+ N_{\rm gauge}= \frac{N_{O3}}{4}+\frac{\chi(D_{O7})}{4}\, .
}
It will serve as a  consistency check that this number is indeed an integer.

If $H^2_-({\cal M}) \neq 0$ with  some non-trivial gauge-flux turned on,
one should also check whether the net $D5$-charge vanishes, i.e.
\eq{
\sum_a \int_{\cal M} \omega \wedge (\text{tr}{\cal F}_a \wedge D_a + \text{tr}{\cal F}_{a'} \wedge D_{a'}) =0
}
for all $\omega \in H^2_-({\cal M})$.

\subsubsection*{Instanton zero modes}

Next, we discuss the zero mode structure of Euclidian $D3$-brane
(short $E3$ instantons) with special emphasis on the
case of a poly-instanton correction\footnote{For a general
review on D-brane instantons see \cite{Blumenhagen:2009qh}.}.
Thus, we describe the ``T-dual'' of the
the zero mode analysis carried out in \cite{Blumenhagen:2008ji}.
There, instanton $a$ was an Euclidian $E1$-instanton
wrapping a rigid curve of genus zero, i.e. an isolated
$\IP^1$. Instanton $b$, however, was an Euclidian $E1$-instanton
wrapping a rigid curve of genus one, i.e. a torus.
The single complex Wilson line Goldstinos were just the
right zero modes to generate an instanton correction
to the gauge kinetic function, respectively, the instanton
action $S_a$.

Now,  in the case of $\Omega \sigma (-1)^{F_L}$ orientifolds,
a former $E1$ instanton becomes a Euclidian $E3$ instanton
wrapping a divisor ${E}$ on ${\cal M}$, i.e. a complex surface.
In  such a case, the instanton zero modes are still
counted by certain cohomology classes on ${E}$, namely
$H^{n,0}(E)=H^{n}(E,\cO)$, $n=0,1,2$. However, one
has to distinguish between $\sigma$-even and odd classes
$H^{n,0}(E)=H_+^{n,0}(E)+H_-^{n,0}(E)$.
Note that, for such equivariant cohomologies, one defines two chiral indices.
First, there is the usual holomorphic Euler
characteristic of the divisor $E$
\beq
  \chi(E,{\cO}_E)=\sum_{i=0}^2 (-1)^i\, h^i(E,{\cO}_E) =\sum_{i=0}^2 (-1)^i\, h^{i,0}(E) \;
\eeq
and second one can define an index taking the $\mathbb Z_2$ action
of $\sigma$ into account
\beq
\label{lefschetzdef}
  \chi^\sigma(E,{\cO}_E)=\sum_{i=0}^2 (-1)^i\, \left( h_+^{i,0}(E)-h_-^{i,0}(E)\right) \; .
\eeq
The Lefschetz fixed point theorem for our case of interest states that,
if the fixed point set of the involution intersects the divisor $E$
in an isolated curve $M^\sigma=O7\cap E$, then
\eq{
\label{lefschetzone}
     \chi^\sigma(E,{\cO}_E)=-{1\over 4} \int_{M^\sigma} [E]\,,
}
where $[E]\in H^2({\cal M})$ is Poincar\'e dual to the divisor $E$.
For the equivariant Betti numbers a similar theorem applies.
Under the same assumptions, one has
\eq{
\label{lefschetztwo}
      L^\sigma({E})=\sum_{i=0}^4 (-1)^i\,  (b^i_+ - b^i_-)=\chi(M^\sigma)\, .
}
In certain cases, these two index theorems are already sufficient
to uniquely determine all non-vanishing equivariant cohomology classes.
In all the other cases, we employ the algorithm presented in \cite{Blumenhagen:2010ed}
which is based on the formalism presented in \cite{Blumenhagen:2010pv} for
the computation of line bundle valued cohomology classes
over toric varieties\footnote{Each cohomology class
has a representative which is a rationom in the homogeneous
coordinates. Once the action of the involution is fixed,
one can determine its action on the relevant
representatives and then run through the long exact sequences
to finally determine the equivariant cohomology classes.
We have used the {\bf cohomCalg} implementation  to carry
out these computations and in particular to
determine the relevant representatives.}.

The general  zero mode structure for an $E3$-brane instanton
wrapping  a divisor $E$, which is invariant under the involution
$\sigma$, was worked out in detail
in \cite{Blumenhagen:2010ja} (see \cite{Bianchi:2011qh,Grimm:2011dj} for
instantons with additional gauge flux).
For a stack of $N$ such Euclidian branes
of type $O(N)$, the zero mode
spectrum  is summarized in Table \ref{tabledeformb}.

\begin{table}[ht]
  \centering
  \begin{tabular}{c|c|c}
    zero modes                    &  statistics   & number \\
    \hline \hline
                      &    &  \\[-0.4cm]
    $(X_\mu,\theta_\alpha)$       & (bose, fermi) & $H_+^{0,0}(E) \times
    \Ysymm + H_-^{0,0}(E)\times  \Yasymm $ \\[0.1cm]
    $\ov\tau_{\dot\alpha}$        & fermi         & $H_-^{0,0}(E) \times
    \Ysymm + H_+^{0,0}(E)\times  \Yasymm $ \\[0.1cm]
    $\gamma_{\alpha}$             & fermi         & $H_+^{1,0}(E) \times
    \Ysymm + H_-^{1,0}(E)\times  \Yasymm $     \\[0.1cm]
    $(w, \ov\gamma_{\dot\alpha})$ & (bose, fermi) & $H_-^{1,0}(E) \times
    \Ysymm + H_+^{1,0}(E)\times  \Yasymm $ \\[0.1cm]
    $\chi_{\alpha}$               & fermi         & $H_+^{2,0}(E) \times
    \Ysymm + H_-^{2,0}(E)\times  \Yasymm $  \\[0.1cm]
    $(c , \ov\chi_{\dot\alpha} )$ & (bose, fermi) & $H_-^{2,0}(E) \times
    \Ysymm + H_+^{2,0}(E)\times  \Yasymm $
  \end{tabular}
  \caption{\small Zero modes for $O(N)$-instanton. }
  \label{tabledeformb}
\end{table}

\noindent
For an $SP(N)$ instanton, with necessarily $N$ even, $\Yasymm$
and $\Ysymm$ are exchanged.
The zero modes $X_\mu$, $\theta_\alpha$ and $\ov\tau_{\dot\alpha}$ are also
called universal zero modes, as they do not depend on the internal geometry of
the four-cycle $E$. The remaining ones can be considered as  Wilson line and
deformation zero modes, i.e.~as Goldstone bosons and Goldstinos of brane
deformation moduli. In the second column we have indicated whether
the orientifold projection leaves just a fermionic zero mode invariant
or a bosonic and a fermionic zero mode.

Given a conformal field theory, the sign in the M\"obius strip amplitudes
uniquely distinguishes between $O/SP$ instantons.
From that we extract the following purely geometric   characterization
of  $O/SP$ instantons (in case we are only using $O7^-$-planes):
\begin{itemize}
\item Placing an $E3$ instanton right on top of an $O7$-plane
      gives an $SP$-instanton.

\item If the divisor $E$ intersects the $O7$-plane in a curve,
      we have four additional Neumann-Dirichlet type boundary conditions,
     which due to \cite{Gimon:1996rq} changes the former $SP$- to an
     $O$-projection.
     Thus, we have an $O$-instanton.

\item For a divisor $E$ which is parallel to the $O7$ plane, in the sense
     $E\cap O7=\emptyset$, the respective instanton is expected to be $SP$.
\end{itemize}

For contributions to the holomorphic superpotential respectively
the gauge kinetic function, the
anti-holomorphic $\ov\tau^{\dot\alpha}$ zero modes have to be removed. This
happens for a single  instanton being  placed in an
orientifold invariant position
with an $O(1)$  projection, which corresponds to an $SP$-type
projection for the fictitious space-time filling $D7$-branes.
If an instanton wrapping a surface $E$ satisfies this,
we also say ``it is $O(1)$''.
For instanton $a$ there should not be any further zero modes,
i.e. $H^{1,0}(E)=H^{2,0}(E)=0$.
For instanton $b$, the former (purely fermionic)
Wilson line Goldstinos of the $E1$ instanton,
for an $E3$ instanton, can  arise from
either Wilson line or deformation Goldstinos,
counted by $H_+^{1,0}(E)+H_+^{2,0}(E)$.
Therefore, we have the two possibilities listed
in Table \ref{tablepolyzeros} for the
zero mode structure of the poly-instanton correction
to the superpotential.
\begin{table}[ht]
  \centering
  \begin{tabular}{c|c|c}
    class & Instanton $a$  &  Instanton $b$    \\
    \hline \hline
      &    &  \\[-0.4cm]
    $H_+^{0,0}(E)$ &  1  & 1 \\[0.1cm]
    $H_+^{1,0}(E)$ &  0  & $1\ \vert\ 0$ \\[0.1cm]
    $H_+^{2,0}(E)$ &  0  & $0\ \vert\ 1$ \\[0.1cm]
    $H_-^{n,0}(E)$ &  0  & 0
  \end{tabular}
  \caption{\small Zero modes for poly-instantons $a$ and $b$ .}
  \label{tablepolyzeros}
\end{table}

A well known example of a divisor having just one deformation
zero mode is certainly a K3-surface. However, the question
is whether for an $O(1)$-instanton the deformation Goldstino
can be in $H_+^{2,0}(E)$. For the second class of $b$-instantons
one first needs explicit examples of surfaces with a single
complex Wilson line. Second, as above the Wilson line
need to be in $H_+^{1,0}(E)$. Even though a couple of brief
arguments were already given
in \cite{Cicoli:2011yy},  let us analyze these two a priori options
in more detail.

\subsubsection*{Analysis of \boldmath{$H_+^{2,0}(E)$}}

Recall that via contraction with the $\Omega_{3,0}$ form
of the Calabi-Yau threefold, $H^2(E,{\cal O})$ is related to the
sections of the normal
bundle of $E\subset {\cal M}$ as
\eq{  H_\pm^2(E,{\cal O})=H_\mp^0(E,N_E)\, .
}
Note that, due to the adjunction formula, the
normal bundle is equal to the canonical bundle of $E$, i.e.
$N_E=K_E$.

Let us  consider  a K3  divisor $E$
which intersects the $O7$-plane in a curve.
Due to our former characterization, an $E3$-brane wrapping this K3 is $O(1)$.
Applying  the Lefschetz fixed point theorem in eq.\eqref{lefschetzone},
gives $\chi^\sigma(K3,{\cal O})=0$ and therefore
$h^2_-(E,{\cal O})=1$.
Thus, we see that for this $O(1)$ instanton we cannot have
 $h^2_+(E,{\cal O})=1$.

Now, let us assume that $H_+^2(E,{\cal O})=1$.
This corresponds to a section of the normal bundle of $E$, which
changes sign under $\sigma$.
Intuitively this means that the surface $E$ is ``parallel''
to the $O7$ plane and therefore supports an $SP$-instanton,
which will not
contribute to the superpotential (at least not without
invoking new mechanism to soak up the extra zero modes).

To conclude, for an $O(1)$ instanton wrapping a K3-surface,
the latter cannot be in  $H_+^{2,0}(E)$. Thus,
the Hodge-diamond splits as :
\begin{table}[ht]
  \centering
  \begin{tabular}{ccccc}
    & & $1_+ $ & & \\
   & 0 & & 0 & \\
  $1_-$ & & $\!\! \!\! h_+^{11}+h_-^{11}\!\! \!\! $ & & $1_-$ \\
   & 0 & & 0 & \\
    & & $1_+$ & & \\
  \end{tabular}
\end{table}

\subsubsection*{Analysis of \boldmath{$H_+^{1,0}(E)$}}

Now consider the case that the divisor admits a single
complex Wilson line Goldstino so that
the Hodge diamond of the divisor $W$ has the form:

\begin{table}[ht]
  \centering
  \begin{tabular}{ccccc}
    & & 1 & & \\
   & 1 & & 1 & \\
  0 & & $\!\!  h^{11}\!\!  $ & & 0 \\
   & 1 & & 1 & \\
    & & 1 & & \\
  \end{tabular}
\end{table}

\noindent
A class of examples of such surfaces $W$ are $\IP^1$ fibrations
over tori\footnote{
It was proposed that a $T^4$-divisor  is a   potential
candidate\cite{Cicoli:2011yy,Cicoli:2011ct}.
Clearly, such a divisor  has more zero modes
as
\eq{
(h^{00}(T^4),h^{10}(T^4),h^{20}(T^4),h^{11}(T^4))=(1,2,1,4)\, .
}
and one has to make the deformation and one Wilson line
Goldstino massive via for instance turning on fluxes.
For the Wilson line Goldstino a pure gauge flux does not work
\cite{Bianchi:2011qh}.},
which, in the following, we also call $W$-surfaces.
A toric realization  for a simple such fibration is 

\begin{table}[ht]
  \centering
  \begin{tabular}{c|ccccc}
     & $x_1$  & $x_2$  & $x_3$  & $x_4$  & $x_5$        \\
    \hline
    0 & 1 & 1 & 0 & 0 & 0 \\
    3 & 0 &-1 & 1 & 1 & 1 \\
  \end{tabular}
\end{table}

\noindent
We define the divisor  $F=\{x_5=0\}=\IP^1\sqcup\IP^1\sqcup\IP^1$ and 
the base is given by $\Sigma=\{x_2=0\}$, which is the torus $\IP_{111}[3]$.
One indeed finds the Hodge diamond above with $h^{11}=2$.
Moreover $c_1(W)=2\, \Sigma + F$ so that $W$ is non-spin.
In contrast to the deformations, there is no immediate argument why the Wilson line should not
be in $H^1_+(W,\cO)$.

Let us make one comment motivated by  the models discussed
in \cite{Cicoli:2011yy,Cicoli:2011ct}.
Such a  $W$-divisor is  not a Calabi-Yau twofold so that,
opposed to $K3$ or $T^4$,  it cannot be fibered over a base to give a
Calabi-Yau threefold. However, the possibility of generating poly-instanton
effects with the $K3$ or $T^4$ fibered geometries used in
\cite{Cicoli:2011yy,Cicoli:2011ct} is not
entirely ruled out in a more involved setup with fluxes.

\section{Orientifolds with poly-instantons}

As a proof of principle, we
now present concrete examples of Type IIB orientifolds
on compact Calabi-Yau threefolds which contain
such $W$-surfaces as toric divisors.
This includes that we explicitly identify
orientifold projections so that
the divisor $W$  is $O(1)$ with $H^1_+(W,\cO)=1$.
With future applications in the framework of the
LARGE volume scenario in mind, we focus on threefolds which also
have shrinkable del Pezzo surfaces, thus featuring a
swiss-cheese type K\"ahler potential.
We are here applying the rules for Type IIB orientifold
model building laid out in \cite{Blumenhagen:2008zz,Collinucci:2008sq,Cicoli:2011it,Cicoli:2011qg}, to which
we refer for more information.

As a word of warning:
we will employ  the sufficient
conditions for certain instanton corrections laid out in the last
section. This means that the superpotentials we write down might  not
be complete.

\subsection{Example A}

In this subsection we present a concrete Type IIB orientifold
on a Calabi-Yau manifold that not just admits a divisor
$W$ but in addition also two rigid and shrinkable del Pezzo
divisors.

The threefold  is defined by the following toric data
\begin{table}[ht]
  \centering
  \begin{tabular}{c|cccccccc}
     & $x_1$  & $x_2$  & $x_3$  & $x_4$  & $x_5$ & $x_6$  & $x_7$ & $x_8$        \\
    \hline
    2  & -1 & 0 & 1 & 1 & 0 & 0 &  0 & 1  \\
    4  & -2 & 0 & 2 & 2 & 1 & 0 &  1 & 0  \\
    2  & -3 & 0 & 2 & 1 & 1 & 1 &  0 & 0  \\
    2  & 1 & 1 & 0 & 0 & 0 & 0 &  0 & 0  \\
  \end{tabular}
 \end{table}

\noindent
and has Hodge numbers $(h^{21}, h^{11}) = (72, 4)$ with Euler number $\chi=-136$. The Stanley-Reisner ideal reads
\begin{equation}
{\rm SR}=  \left\{x_1\,x_2,  x_4\,x_7,  x_5\,x_7,  x_1\,x_4\,x_8,
x_2\,x_5\,x_6,  x_3\,x_4\,x_8,  x_3\,x_5\,x_6,  x_3\,x_6\,x_8 \right\}\, .
\end{equation}
The intersection form is most conveniently displayed choosing the basis of smooth divisors as $\{D_1, D_6, D_7, D_8\}$.
Then, the triple intersections on the Calabi-Yau threefold have the form
\eq{
\label{intersect718}
I_3=9D_1^3-3D_1^2D_6-4D_6D_8^2+&D_1D_6^2-3D_6^3+2D_7D_8^2+2D_6D_7D_8\, \\
&\ +2D_6^2D_7-2D_7^2D_8-2D_6D_7^2+2D_7^3\, .
}
Writing the K\"ahler form in the above basis of divisors as
$J=t_1D_1+t_6D_6+t_7D_7+t_8D_8$, the resulting volume form in terms of
two-cycle volumes $t_i$ takes the form
\bea
& & {\cal V}\equiv \frac{1}{3 !} \int_{\cal M} J \wedge J\wedge J  =\frac{1}{6}\Bigl(9 t_1^3 - 9 t_1^2 t_6 + 3 t_1 t_6^2 - 3 t_6^3 + 6 t_6^2 t_7 - 6 t_6 t_7^2 + 2 t_7^3 \nonumber\\
& & \hskip 1.6in + 12 t_6 t_7 t_8 - 6 t_7^2 t_8 - 12 t_6 t_8^2 + 6 t_7
t_8^2\Bigr)\, .
\eea
Expanding the K\"ahler form as $J=r^i [K_i]$, for
 the following four divisors  $\{K_i, i=1,\dots,4\}$
the K\"ahler cone is given simply by $r^i>0$:
\eq{
\label{Kaehler718}
K_1 &=2D_6+2D_7+D_8,\, \\
K_2 &=2D_6+3D_7+D_8,\,\\
K_3 &=D_1+4D_6+4D_7+2D_8,\, \\
K_4 &=D_1+5D_6+5D_7+2D_8\, .
}
For the K\"ahler parameters $t_i$
this translates into
\begin{equation}
\label{Kaehlercone718}
 -t_6+t_7>0,\ -2t_1+t_6-t_7+t_8>0,\ t_1-t_6+2t_8>0,\ t_6-2t_8>0\, .
\end{equation}

\noindent
Defining the four-cycle volumes
\begin{equation}
 \tau_i=\frac{1}{2} \int_{D_i} J \wedge J\, ,
\end{equation}
we find
\eq{
 \tau_1 &=\textstyle{ \frac{1}{2}} (-3t_1+t_6)^2,\, \\
 \tau_6 &=-\textstyle{ \frac{3}{2}}t_1^2+t_1t_6-\textstyle{\frac{3}{2}} t_6^2
   +2 t_6t_7-t_7^2+2t_7t_8-2t_8^2,\, \\
\tau_7 &=(t_6-t_7+t_8)^2,\, \\
\tau_8 &=(2t_6-t_7)(t_7-2t_8)\, .
}
Taking into account the K\"ahler cone constraints \eqref{Kaehlercone718},
the volume can be written in the strong swiss-cheese form
\eq{
\label{volumeA}
{\cal V}&=\textstyle{ \frac{1}{9}}\Bigl(\frac{1}{\sqrt 2}
(\tau_1+3\tau_6+6\tau_7+3\tau_8)^{3/2}-\sqrt{2}\tau_1^{3/2}-3\tau_7^{3/2}-3(\tau_7+\tau_8)^{3/2}\Bigr)\,.
}
The above volume form shows that the large volume limit is given by
 $\tau_6 \rightarrow \infty$ while keeping the other four-cycles small.

Computing the Hodge diamonds, one finds that the divisor
$D_1$ is a $\IP^2$ surface, $D_7$ is a $dP_7$ surface and the
divisor $D_8$ is indeed the desired  Wilson line divisor $W$.
Moreover, the strong swiss-cheese form of the volume implies that both the
$\IP^2$ and the $dP_7$ divisor are  shrinkable to a point in  ${\cal M}$.

From the volume \eqref{volumeA} it seems that the
divisor $D_7+D_8$ might also be a
del Pezzo surface, but, as apparent from the toric data,  the only
monomial of this degree
is $x_7\,x_8$. Therefore, this defines  a singular surface which
is just the intersection of the $dP_7$ and the $W$ divisors.
It can be  shown  that these two divisors intersect over
a genus one curve in the Calabi-Yau threefold.
One can also show that the intersection $D_5\cap D_8$
in the Calabi-Yau threefold is also a $T^2$ curve, while $D_1$ intersects
$D_5$ in a genus zero curve and does not intersect the  divisors
$D_7$ and $W$.
See also Table \ref{tabledivsA}.

\subsubsection*{Orientifold projections}

Next, we have to specify an orientifold projection so that
$W$ is $O(1)$ and that the
Wilson line Goldstino is in $H_{+}^1(W,\cO)$. Restricting
to the case that we just flip the sign of one homogeneous
coordinate, we find two inequivalent  involutions
$\sigma:$ $\{x_7 \leftrightarrow -x_7, x_4 \leftrightarrow -x_4 \}$,
which have  $h^{11}_-({\cal M})=0$.

\vspace{0.4cm}
\noindent
{\it Involution $\sigma: x_7 \leftrightarrow -x_7$}

\vspace{0.2cm}
Let us  discuss the involution  $\sigma: x_7 \leftrightarrow -x_7$
in more detail. Taking also the constraints from the
Stanley-Reisner ideal into account,
the fixed point set of the toric four-fold is
\eq{
\{{\rm Fixed}_{x_7 \leftrightarrow -x_7}\}=\{D_5\,,\, D_7\,,\, D_1D_4D_6\,,\,
D_2D_4D_6\}\, .
}
This fixed point set intersects the respective $\sigma$-invariant
hypersurface(see below) so that we get a number of $O7$ and $O3$-planes.
Concretely, there are two $O7$-components
\eq{
O7=D_5 \sqcup D_7 \, .
}
In order to determine the $O3$-planes, using the intersection form
\eqref{intersect718}, we  can  compute the
triple intersection numbers as
\eq{
        D_1\, D_4\, D_6 = 1, \qquad  D_2\, D_4\, D_6=4
}
so that  in total there are five $O3$-planes.
As mentioned,  the $O7$-component $D_7$ is a $dP_7$ and the independent
Hodge numbers of the $D_5$ are given by
\eq{
(h^{00}(D_5),h^{10}(D_5),h^{20}(D_5),h^{11}(D_5))=(1,0,1,21)\,
}
with Euler number $\chi{(D_5)}=25$.

Applying the Lefschetz theorems \eqref{lefschetzone}
and \eqref{lefschetztwo}, we find that the
Hodge numbers of the $W$-divisor $D_8$ are
\eq{
(h^{00}(W),h^{10}(W),h^{20}(W),h^{11}(W))=(1_+,1_+,0,2_+)\, ,
}
which is what we want for poly-instanton corrections.
The divisor $D_1=\IP^2$ has the Hodge numbers
\eq{
(h^{00}(D_1),h^{10}(D_1),h^{20}(D_1),h^{11}(D_1))=(1_+,0,0,1_+)\, .
}
Moreover, the relative positions of $W$, $D_1$
and  the two components of the $O7$-plane are shown in Figure 1.

\vspace{0.3cm}

\begin{figure}[ht]
  \centering
  \includegraphics[width=0.46\textwidth]{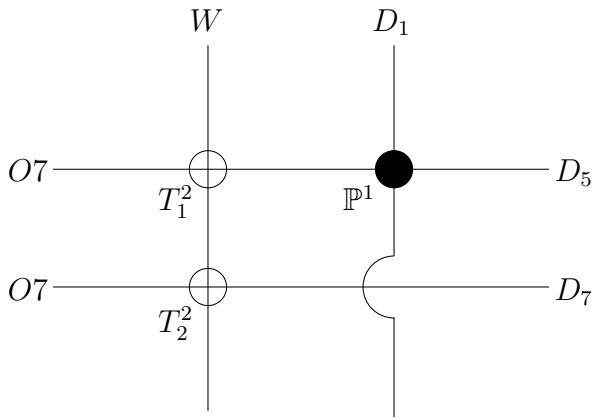}
 \begin{picture}(0,0)
   \put(-141,147){$W$}
   \put(-72,147){$D_1$}
   \put(-210,90){$O7$}
   \put(-210,45){$O7$}
   \put(-3,90){$D_5$}
   \put(-3,45){$D_7$}
    \put(-153,79){$T^2_1$}
    \put(-153,33){$T^2_2$}
    \put(-83,79){$\IP^1$}
 \end{picture}
  \caption{\small Relative positions of $O7$-planes and divisor $W$ for example A with involution $\sigma: x_7 \leftrightarrow -x_7$.}
  \label{fig:oplaneA}
\end{figure}

The simplest solution to the $D7$-brane tadpole cancellation condition is that we place eight $D7$-branes right on top of the $O7$-plane.
We cancel the Freed-Witten anomalies for branes on the divisors
$D_1,D_7$ and $D_8$ by choosing ${\cal F}_{1}={\cal F}_{7}={\cal F}_{8}=0$ and
turning on a  global half-integer quantized $B$ field
with $c_1(B)={1\over 2}( D_1+D_7+D_8)$.
Using \eqref{d3tadsimple} the contribution to the $D3$-brane tadpole is
\eq{
N_{D3}+\frac{N_{\text{flux}}}{2}+N_{\rm gauge}=\frac{N_{O3}}{4}+\frac{\chi(D_{O7})}{4}
=\frac{5+(10+25)}{4}=10\, .
}

In this case, we will get two contributions to the non-perturbative
superpotential. First, there will be an $E3$-instanton
wrapping the divisor $D_1=\IP^2$ and second, the ${\cal N}=1$ super Yang-Mills gauge theory on the
$dP_7$ divisor develops a gaugino condensate.
Thus, one gets a non-perturbative superpotential of the form
\eq{
\label{superone}
   W=A_1\, \exp(-2\pi T_1) + A_7\, \exp(-a_7 T_7) \, ,
}
where the $T_i$'s are the complexified K\"ahler moduli
appearing in the  ${\cal N}=1$ chiral supermultiplet.
It is  defined as
$T_i = \tau_i + i\rho_i$ for a holomorphic, isometric involution such that
$h^{1,1}_{-}({\cal M}) =0$, where
the $\rho_i$ denote the components of the $C_4$ axion.
The divisor $W$ seems to have   the right zero modes to generate
a poly-instanton correction to this.
However, there is one subtlety with possible charged matter zero modes appearing
on the intersection curve $C=E3\cap D7$, i.e. localized at $D_1 \cap D_5 =
\IP^1$ and  $D_{5,7} \cap D_8 = T^2$ in our case.

Since there is no non-trivial line
bundle carried by the $D7$-branes, the number of such matter zero modes is counted by the cohomology groups
$H^i(C, K^{1/2}_C)$,  where $i=\{0,1\}$ and $K^{1/2}_C$ is the spin-bundle of $C$.
Since $D_1\cap D_5=\IP^1$ and $H^*(\IP^1,{\cal O}(-1))=(0,0)$, there will
be no extra matter zero modes.
Since $W$ intersects the  $SO(8)\times SO(8)$ stacks of $D7$ branes over
a  $T^2$ and $H^*(T^2, {\cal O})=(1,1)$, there appear
extra vector-like zero modes. As discussed in \cite{Blumenhagen:2008zz},
these zero mode can pair up and become massive, if one has
a non-trivial Wilson line on $T^2$.
For this purpose, one must have the freedom to turn on
an additional gauge bundle on  the $D7$-brane divisor $D$, whose
restriction on the curve $C=T^2$ is a non-trivial Wilson line.
Therefore, an additional gauge bundle $R$ which is supported
only on two-cycles $C_i\subset D$ that are topological trivial in ${\cal M}$ but do intersect with the curve $C$,
allows one to avoid these extra zero modes.

In our case, both $D_5$ and $D_7$ have more two-cycles
than the ambient Calabi-Yau space so that there
must exist such trivial two-cycles.
Since the $D7$ branes lie right on top of the
$O7$-planes, all two cycles in  $H^{11}(D_7)$ and $H^{11}(D_5)$
are invariant. Therefore, turning on diagonal $U(1)$ gauge flux along the
trivial 2-cycles breaks
the gauge symmetry from $SO(8)\times SO(8)$ to
$U(4)\times U(4)$.
This shows that  the divisor $W$ indeed generates a poly-instanton
correction to \eqref{superone} of the form
\eq{
   W&= A_1\, \exp\left( -2\pi T_1\right) + A_1\, A_8\, \exp\left(
       -2\pi T_1-2\pi T_8\right)+\\
&\phantom{aaaaaaaaaaa}A_7\, \exp\left( -a_7 T_7\right) + A_7\, A_8\, \exp\left(
       -a_7 T_7-2\pi T_8\right)+\ldots\; .
}

This simple example serves as a proof of principle that: a.) the divisor
$W$ can indeed by embedded into a compact Calabi-Yau threefold and
b.) an orientifold projection can be identified so that it is $O(1)$
and $H^1_+(W,\cO)=1$.

\vspace{0.2cm}

To summarize, in our geometry we find divisors with the topological data
listed in Table \ref{tabledivsA}.

\begin{table}[ht]
  \centering
  \begin{tabular}{c|c|c}
    divisor & $(h^{00},h^{10},h^{20},h^{11})$  &  intersection curve    \\
    \hline \hline
      &    &  \\[-0.4cm]
    $D_7=dP_7$ & $(1_+,0,0,8_+)$ & $W: C_{g=1}$ \\
    $D_5$ & $(1_+,0,1_+,21_+)$ & $W: C_{g=1}$ \\
    \hline
    $D_8=W$ & $(1_+,1_+,0,2_+)$ & $D_5: C_{g=1}, \ \ D_7:  C_{g=1}$ \\
    $D_1=\IP^2$ & $(1_+,0,0,1_+)$ & $D_5: C_{g=0}$
  \end{tabular}
  \caption{\small Divisors and their equivariant cohomology. The first two
    line are $O7$-plane components  and the second two divisors
    can support $E3$ instantons. The $D_7$ divisor also supports
    gaugino condensation.}
  \label{tabledivsA}
\end{table}

\vspace{0.4cm}
\noindent
{\it Involution $\sigma: x_4 \leftrightarrow -x_4$}

\vspace{0.2cm}

Let us briefly mention what happens for the second involution
$x_4 \leftrightarrow -x_4$.
Here, the fixed point set reads
\eq{
  \{{\rm Fixed}_{x_4 \leftrightarrow -x_4}\}=
 \bigl\{ D_4,\, D_1\,D_3\,D_8\, ,\, D_1\,D_5\,D_6\, ,\, &D_1\,D_6\,D_7\,,\\
        &D_2\,D_3\, D_8\, ,\, D_2\, D_6\,D_7 \bigr\}\, .
}
Therefore, we have a single  $O7$-plane component on $D_4$
with $\chi{(D_4)}=35$, while there
are five $O3$-planes with one located at  $D_1\cap{D_5}\cap{D_6}$ and
four located at
$D_2\cap{D_6}\cap{D_7}$. The contribution to the $D3$-brane tadpole
in this case is $\frac{\chi{(D_{O7})}}{4}+\frac{N_{O3}}{4}=\frac{35+5}{4}=10$.

We can employ {\bf cohomCalg} in order to calculate the equivariant cohomology of the divisors. The relevant ones and their  topological data are summarized in
Table \ref{tabledivsA2}.

\begin{table}[ht]
  \centering
  \begin{tabular}{c|c|c}
    divisor & $(h^{00},h^{10},h^{20},h^{11})$  &  intersection curve    \\
    \hline \hline
      &    &  \\[-0.4cm]
    $D_4$ & $(1_+,0,2_+,29_+)$ & $D_1: C_{g=0}$ \\
    \hline
    $D_7=dP_7$ & $(1_+,0,0,5_++3_-)$ & $W: C_{g=1}$ \\
    $D_8=W$ & $(1_+,1_+,0,2_+)$ & $ D_7:  C_{g=1}$ \\
    $D_1=\IP^2$ & $(1_+,0,0,1_+)$ & $D_4: C_{g=0}$
  \end{tabular}
  \caption{\small Divisors and their equivariant cohomology. The first
    line is an  $O7$-plane   and the last three divisors
    support $E3$ instantons.}
  \label{tabledivsA2}
\end{table}

\begin{figure}[ht]
  \centering
  \includegraphics[width=0.46\textwidth]{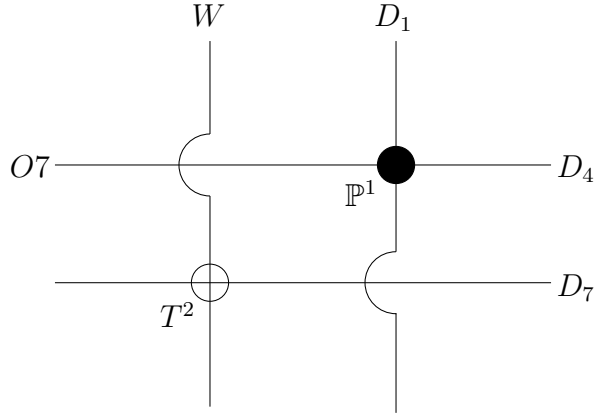}
 \begin{picture}(0,0)
   \put(-141,147){$W$}
   \put(-72,147){$D_1$}
   \put(-210,90){$O7$}
   \put(-3,90){$D_4$}
   \put(-3,45){$D_7$}
    \put(-153,33){$T^2$}
    \put(-83,79){$\IP^1$}
 \end{picture}
  \caption{\small Relative position of $O7$-planes and divisor $W$ for example A with involution $\sigma: x_4 \leftrightarrow -x_4$.}
  \label{fig:oplanesA2}
\end{figure}

\noindent
The relative positions of these divisors are shown in figure
\ref{fig:oplanesA2}.

\vspace{0.3cm}

\noindent
Since the $O7$-plane and the accompanying stack of eight $D7$-branes
only intersects $D_1$ in a genus zero curve $\IP^1$,
there are  no extra vector-like zero modes. Therefore, the gauge symmetry
is $SO(8)$.

For this model we can identify two kinds of non-perturbative
contributions to the superpotential and their poly-instanton
corrections.
First, an $E3$ instanton wrapping the surface $D_1=\IP^2$ contributes
to the superpotential and a second $E3$ instanton wrapping
$W$ generates a poly-instanton correction to that.
Second, an $E3$-instanton on $D_7=dP_7$ contributes to the superpotential.
Since, $h^{11}_-(D_7)=3$, there can be gauge fluxes in
$F\in H^{11}_-(D_7)$ which, as mentioned in section 2,
do not spoil the $O(1)$-property.
Therefore, we have a sum over these fluxed instantons.
Moreover, in general these fluxes will also make the
vector-like zero modes on the $T^2=W\cap D_7$ massive
so that the fluxed instantons also receive a poly-instanton
contribution.

From these considerations, for this model we  expect a rather
intricate total superpotential with leading order terms
\eq{
W= &A_1\, \exp\left( -2\pi T_1\right) + A_1\, A_8\, \exp\left(
       -2\pi T_1- 2\pi T_8\right) +
 A_7\, \exp\left( -2\pi T_7\right) +\\
  &\!\!\!\!\!\!\!\!\!\sum_{F_i\in H^{11}_-(D_7)}\!\!\!\!\!
      A^{(i)}_7\exp\left( -2\pi T_7 - f_i \tau \right)
     +A^{(i)}_7\, A_8 \exp\left( -2\pi T_7 - f_i \tau
     -2\pi T_8 \right)\ldots
}
where $\tau$ denotes the axio-dilaton superfield and
$f_i\simeq \int_{D_7} F^i\wedge F^i$ depends on the gauge flux.

\subsection{Example B}

In our scan for Calabi-threefolds having the desired divisors,
we also found an example for which not all K\"ahler deformations
are torically realized.
The Calabi-Yau ${\cal M}$ is given by a hypersurface in a toric variety
with defining data

\begin{table}[ht]
  \centering
  \begin{tabular}{c|cccccccc}
     & $x_1$  & $x_2$  & $x_3$  & $x_4$  & $x_5$ & $x_6$  & $x_7$ & $x_8$        \\
    \hline
    3  & 1 & 0 & 0 & 0 & 0 & 0 &  0 & 1  \\
    4  & 1 & 1 & 1 & 0 & 0 & 0 &  1 & 0  \\
    12 & 4 & 3 & 3 & 0 & 1 & 1 &  0 & 0  \\
    6  & 2 & 2 & 1 & 1 & 0 & 0 &  0 & 0  \\
  \end{tabular}
\end{table}

\noindent
with Hodge numbers $(h^{21}, h^{11}) = (89, 5)$ and Euler number $\chi=-168$.
Since $h^{11}=5$ exceeds  the number of toric equivalence  relations,
one K\"ahler deformation is non-toric.
The Stanley-Reisner ideal is
\eq{
{\rm SR}=  \bigl\{x_3\,x_4\,,\,  x_3\,x_7\,,\,  x_7\,x_8\,,\,
x_1\,x_2\,x_4\,,\,  &x_1\,x_2\,x_8\,,\,  x_1\,x_5\,x_6\,,\\
   &  x_2\,x_4\,x_7 \,,\,
  x_3\,x_5\,x_6\,,\,  x_5\,x_6\,x_8 \bigr\}\, .
}
The triple intersection form in the basis of smooth divisors $\{D_4, D_6, D_7,
D_8\}$ reads
\eq{
\label{intersect1004}
I_3=9D_4^3-9D_4^2D_8+9D_4D_8^2&-3D_4^2D_6+3D_4D_6D_8-6D_6D_8^2 \\
&+D_4D_6^2-D_6^3+2D_6^2D_7-6D_6D_7^2+18D_7^3\, .
}
Writing the K\"ahler form in the above basis of divisors as $J=t_4D_4+t_6D_6+t_7D_7+t_8D_8$, the resulting volume form in terms of two-cycle volumes $t_i$ is given as,
\eq{
&  {\cal V}\equiv \frac{1}{3 !} \int_{\cal M} J \wedge J\wedge J  =\frac{1}{6}\Bigl(9t_4^3 -t_6^3+6t_6^2 t_7 +18 t_7^3 -9 t_4^2(t_6 + 3 t_8)\\
&  \hskip2in +3 t_4(t_6+3t_8)^2 -18 t_6(t_7^2+t_8^2)\Bigr)\, .
}
There are four generators $\{K_i, i=1,2,3,4\}$ for the toric
K\"ahler cone. We can expand the K\"ahler form  as $J=r^i [K_i]$ with $r^i>0$
and
\eq{
\label{Kaehler1004}
K_1 &=2D_4+4D_6+D_7+D_8\, , \\
K_2 &=D_4+3D_6+D_7\, ,\\
K_3 &=6D_4+12D_6+4D_7+3D_8\, ,\\
K_4 &=2D_4+6D_6+2D_7+D_8\, .
}
In the basis $J=t_4D_4+t_6D_6+t_7D_7+t_8D_8$ the K\"ahler cone
is given by
\begin{equation}
\label{Kaehlercone1004}
 t_6-3t_7>0,\ t_4-2t_8>0,\ t_4-t_6+2t_7>0,\ -3t_4+t_6+2t_8>0\, .
\end{equation}
For the corresponding four-cycle volumes we find
\eq{
 \tau_4 &=\frac{1}{2} (-3t_4+t_6+3t_8)^2\, ,\\
 \tau_6 &=-\frac{3}{2}t_4^2-\frac{1}{2}t_6^2+2t_6t_7+t_4(t_6+3t_8)-3(t_7^2+t_8^2)\, ,\\
\tau_7 &=(t_6-3t_7)^2\, ,\\
\tau_8 &=-\frac{3}{2}(3t_4-2t_6)(t_4-2t_8)\, .
}
Taking into account  the K\"ahler cone constraints
\eqref{Kaehlercone1004},  the volume can be written again in the
strong swiss-cheese form
\eq{
\label{volumeB}
{\cal
  V}=\frac{1}{9}\Bigl(\sqrt{2}(2\tau_4+3\tau_6+\tau_7+\tau_8)^{3/2}-\sqrt{2}\tau_4^{3/2}-\tau_7^{3/2}-\sqrt{2}(\tau_4+\tau_8)^{3/2}\Bigr)\, ,
}
which shows that the large volume limit is defined as
$\tau_6 \rightarrow \infty$ while keeping the other four-cycles small.

Computing the Hodge diamonds, one finds that the divisor $D_4$ is a
$\IP^2$ and the divisor $D_8$  a Wilson line divisor.
For the Hodge diamond of the $D_7$ divisor {\bf cohomCalg}
gives the output
$(h^{00}(D_7),h^{10}(D_7),h^{20}(D_7),h^{11}(D_7))=(2,0,0,2)$.
Therefore, the locus $x_7=0$ seems to have two $\IP^2$ components,
$D'_7$ and $D''_7$,
of which one linear combination is toric and the other related to the fact that
we have one non-toric element in $H^{11}({\cal M})$.
This suggests that the complete volume form is given by
\eqref{volumeB} with the simple substitution
$\tau_7^{3/2}= (\tau'_7)^{3/2} + (\tau''_7)^{3/2}$.

Again, the
divisor $D_4+D_8$ has  no smooth surface representing it and
the intersection $D_4\cap D_8$
is a  $T^2$ curve. For some purposes below, it can be shown
that the intersection $D_3 \cap D_8$ on the Calabi-Yau threefold is also
a $T^2$ curve. See also Table \ref{tabledivsB}.

\subsubsection*{Orientifold projections}

We identified two orientifold projection
$\sigma:$ $\{x_4 \leftrightarrow -x_4, x_2 \leftrightarrow -x_2\}$
so that the Wilson line Goldstino is in $H_{+}^1(W,\cO)$ and $W$ supports
${\cal O}(1)$ instanton.
For both involutions, one can determine that $h^{11}({\cal M})=4_++1_-$.
Hence the ${\cal N} =1$ K\"ahler coordinates $T_\alpha$
will be modified by the presence of the single odd-modulus
\eq{
G=\int_{D_-} B_2 + i\int_{D_-} C_2\, ,
}
where $D_-\in H^{11}_-({\cal M})$.

\vspace{0.4cm}
\noindent
{\it Involution $\sigma: x_4 \leftrightarrow -x_4$}

\vspace{0.2cm}

Under this involution the fixed point set is
\eq{
\{{\rm Fixed}_{x_4 \leftrightarrow -x_4}\}=\{D_3\, ,\, D_4\,,\, D_1D_2D_7\,,\,
D_2D_5D_6\}\, .
}
Thus, there are  the two $O7$-components
\eq{
O7=D_3 \sqcup D_4 .
}
With the help of the intersection form \eqref{intersect1004}, we find
that $D_1\cap{D_2}\cap{D_7}$ does not  intersect
the hypersurface, while there is one $O3$-plane  on
$D_2\cap{D_5}\cap{D_6}$.
For $D7$-brane tadpole cancellation, we again place eight
$D7$-branes right on top of the $O7$-plane.
The contribution to the $D3$-brane tadpole becomes
\eq{
\frac{\chi(D_{O7})}{4}+\frac{N_{O3}}{4}=\frac{(60+3)+1}{4}=16\,.
}
Since there is no non-trivial gauge field configuration on these
$D7$-branes(see below), the net $D5$-brane charge vanishes too.
For the relevant divisors, we find  the topological data
listed in Table \ref{tabledivsB}.


\begin{table}[h]
  \centering
  \begin{tabular}{c|c|c}
    divisor & $(h^{00},h^{10},h^{20},h^{11})$  &  intersection curve    \\
    \hline \hline
      &    &  \\[-0.4cm]

    $D_3$ & $(1_+,0,4_+,50_+)$ & $W: C_{g=1}$ \\
    $D_4=\IP^2$ & $(1_+,0,0,1_+)$ & $W: C_{g=1}$ \\
    \hline
    $D_7=\IP^2\sqcup \IP^2$ & $(1_++1_-,0,0,1_++1_-)$ &  null  \\
    $D_8=W$ & $(1_+,1_+,0,2_+)$ & $D_3: C_{g=1}, \ \ D_4:  C_{g=1}$ \\
  \end{tabular}
  \caption{\small Divisors and their equivariant cohomology. The first two
    line are $O7$-plane components  and the last two divisors
    support $E3$ instantons. The $D_4$ divisor also supports gaugino condensation.}
  \label{tabledivsB}
\end{table}

\begin{figure}[ht]
  \centering
  \includegraphics[width=0.46\textwidth]{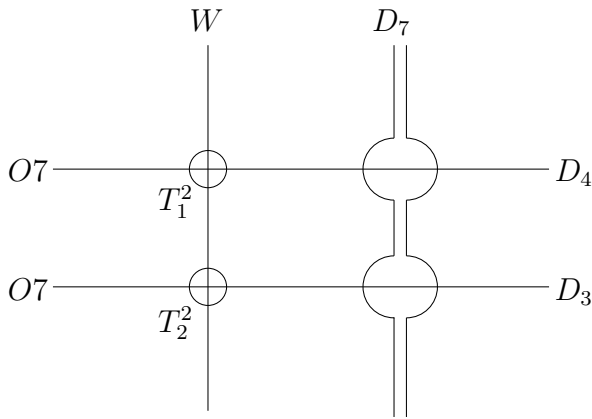}
 \begin{picture}(0,0)
   \put(-141,147){$W$}
   \put(-72,147){$D_7$}
   \put(-210,90){$O7$}
   \put(-210,45){$O7$}
   \put(-3,90){$D_4$}
   \put(-3,45){$D_3$}
    \put(-153,79){$T^2_1$}
    \put(-153,33){$T^2_2$}
 \end{picture}
  \caption{\small Relative position of $O7$-planes and divisor $W$ for example B with involution $\sigma: x_4 \leftrightarrow -x_4$.}
  \label{fig:oplanesB1}
\end{figure}

Note that the equivariant cohomology for $D_7$ suggests that
its two $\IP^2$ components are interchanged by the orientifold
projection. For such a $\sigma$ to be a symmetry one in particular
needs $\tau'_7=\tau''_7$.
The positions of these divisors are also shown in figure
\ref{fig:oplanesB1}.


\noindent

This model has some features which differ from example A.
First, since $D_4=\IP^2$ one does not have the degree of freedom
to give a mass to the vector-like matter zero modes
on the $T^2=W\cap D_4$ intersection. Second, under $\sigma$
the two $\IP^2$ components of $D_7$ are interchanged so that this is
not an $O(1)$ instanton (but a $U(1)$ instanton with a non-zero number
of $\ov\tau_{\dot\alpha}$ zero modes).

Even though  the superpotential receives a contribution from the
gaugino condensation of the pure $SO(8)$ super Yang-Mills theory
on the divisor $D_4$,
due to the vector-like matter zero modes, the $W$ divisor has too many
zero modes to generate an additional  poly-instanton contribution.
Although  the superpotential has the simple
form $W= A_4\, \exp\left( -a_4\, T_4\right)$,
we think that the  presentation of this model was
nevertheless useful for illustrative purposes.

\vspace{0.5cm}
\noindent
{\it Involution $\sigma: x_2 \leftrightarrow -x_2$}

\vspace{0.2cm}

For this involution the
 fixed point set consistent with the SR-ideal is
\eq{
\{{\rm Fixed}_{x_2 \leftrightarrow -x_2}\}=\{D_2\, ,\,  D_1D_4D_7\,,\,
D_4D_5D_6\}\, .
}
The $O7$-plane is located on the divisor  $D_2$
with $\chi{(D_2)}$=63. In addition  there is one $O3$-plane given
by $D_4\cap{D_5}\cap{D_6}$.
Placing  eight $D7$-branes on top of the $O7$-plane,
the contribution to the $D3$-brane tadpole is
\eq{
\frac{\chi{(D_{O7})}}{4}+\frac{N_{O3}}{4}=\frac{63+1}{4}=16\, .
}
The topological data of the interesting divisors are  summarized in
Table \ref{tabledivsB2}.

\newpage
\begin{table}[ht]
  \centering
  \begin{tabular}{c|c|c}
    divisor & $(h^{00},h^{10},h^{20},h^{11})$  &  intersection curve    \\
    \hline \hline
      &    &  \\[-0.3cm]
    $D_2$ & $(1_+,0,5_+,51_+)$ & null \\
     \hline
    $D_4=\IP^2$ & $(1_+,0,0,1_+)$ & $W: C_{g=1}$ \\
    $D_7=\IP^2\sqcup \IP^2$ & $(1_++1_-,0,0,1_++1_-)$ &  null  \\
    $D_8=W$ & $(1_+,1_+,0,2_+)$ & $D_4: C_{g=1}$ \\
  \end{tabular}
\caption{\small Divisors and their equivariant cohomology. The first 
    line is a $O7$-plane component  and the last three divisors
    support $E3$ instantons.}
  \label{tabledivsB2}
\end{table}

\begin{figure}[ht]
  \centering
  \includegraphics[width=0.46\textwidth]{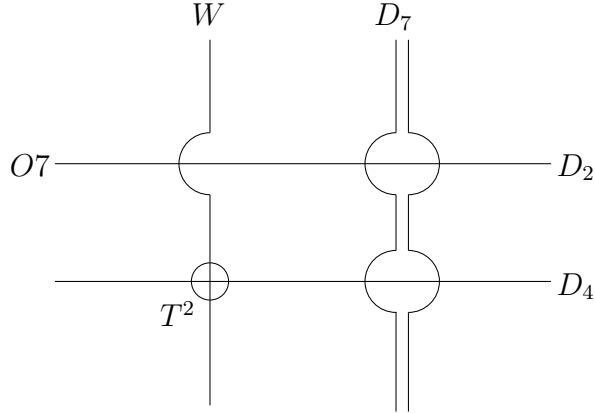}
 \begin{picture}(0,0)
   \put(-141,147){$W$}
   \put(-72,147){$D_7$}
   \put(-210,90){$O7$}
   \put(-3,90){$D_2$}
   \put(-3,45){$D_4$}
    \put(-153,33){$T^2$}
 \end{picture}
  \caption{\small Relative position of $O7$-planes and divisor $W$ for example B with involution $\sigma: x_2 \leftrightarrow -x_2$.}
  \label{fig:oplanesB2}
\end{figure}

\noindent
Their relative position is depicted in figure \ref{fig:oplanesB2}.

\vspace{0.2cm}

\noindent

In this case,
still the pair of $\IP^2$s wrapping $D_7$  support a $U(1)$ instanton
and the $E3$ instanton wrapping $D_4$ does not receive
a  poly-instanton correction, as the vector-like zero modes
on $W\cap D_4$ cannot be made massive.
Therefore, we have only the simple superpotential
$W=A_4\, \exp\left( -2\pi  T_4\right)$.

\subsection{Example C}

The last example provides a model where components
of the orientifold $O7$-plane are non-generic.
The Calabi-Yau ${\cal M}$ is given by a hypersurface in a toric variety
with defining data
\begin{table}[ht]
  \centering
  \begin{tabular}{c|cccccccc}
     & $x_1$  & $x_2$  & $x_3$  & $x_4$  & $x_5$ & $x_6$  & $x_7$ & $x_8$        \\
    \hline
    2  & 1 & 0 & 0 & 0 & 0 & 0 &  0 & 1  \\
    3  & 1 & 1 & 0 & 0 & 0 & 0 &  1 & 0  \\
    9  & 3 & 3 & 0 & 1 & 1 & 1 &  0 & 0  \\
    4  & 2 & 1 & 1 & 0 & 0 & 0 &  0 & 0  \\
  \end{tabular}
\end{table}

\noindent
and has  Hodge numbers $(h_{21},h_{11})=(112,4)$ and
Euler characteristic $\chi({\cal M})=-216$.
The resulting Stanley-Reisner ideal is
\eq{
 {\rm SR}=  \left\{ x_1\, x_3 , x_1\, x_8 , x_2\, x_3 , x_2\, x_7 , x_7\, x_8 , x_4\, x_5\, x_6
\right\}\, .
}
The intersection form is most conveniently displayed choosing
the basis of smooth divisors as $\{ D_1,D_3,D_7,D_8\}$.
Then,  the triple intersections
on the Calabi-Yau threefold have the form
\eq{
      I_3=27 D_1^3 + 9\, D_7^3 + 9\, D_3^3 - 9\, D_3^2\, D_8 + 9\, D_3\, D_8^2
      \, .
}
Expanding the K\"ahler form as
$J=t_1\, D_1 + t_3\, D_3+ t_7\, D_7+ t_8\, D_8$,
the volume of the Calabi-Yau is given by
\eq{
      {\cal V}\equiv {1\over 3!}\int_{\cal M} J\wedge J\wedge J={\textstyle {1\over 2}}\left(9\, t_1^3 +
      3\, t_7^3 +3\, t_3^3 - 9\, t_3^2\, t_8 + 9\, t_3\, t_8^2\right)\, .
}
There are four generators $\{K_i, i=1,2,3,4\}$ for the toric
K\"ahler cone. We can expand the K\"ahler form  as $J=r^i [K_i]$ with $r^i>0$
and
\eq{
\label{KaehlerconeC1}
K_1 &=D_1-D_3-D_8\, , \\
K_2 &=D_1-2D_3-D_7-D_8\, ,\\
K_3 &=D_1\, ,\\
K_4 &=2D_1-2D_3-D_8\,
}
and the K\"ahler cone is
\eq{
\label{kahlerconeB}
           t_7<0, \quad t_3-2\, t_8>0, \quad t_1+t_3-t_7>0,
           \quad -t_3+ t_7 + t_8>0 \, .
}
The corresponding 4-cycle volumes read
\eq{
       \tau_1={\textstyle {27\over 2}}\, t_1^2, \quad
       \tau_7={\textstyle {9\over 2}}\, t_7^2,\quad
       \tau_3={\textstyle{9\over 2}}\, (t_3-t_8)^2, \qquad
       \tau_8={\textstyle{9\over 2}}\, (2\, t_3\, t_8 - t_3^2)\, .
}
Taking into account that the K\"ahler cone constraints \eqref{kahlerconeB}
imply $t_1>0$, $t_3-t_8<0$ and $t_8<0$, the volume can be written in
the strong swiss-cheese form
\eq{
       {\cal V}={\textstyle {\sqrt 2\over 9}}\left(
           {\textstyle {1\over \sqrt{3}}}\, (\tau_1)^{3\over 2}
                - (\tau_7)^{3\over 2}
                 - (\tau_3)^{3\over 2}
                -(\tau_3+\tau_8)^{3\over 2}\right)\, .
}

Computing the Hodge diamond,
one finds that the divisors $D_3$ and $D_7$ are $\IP^2$ surfaces
and that the divisor $D_8$ is a $W$-surface.
The strong swiss-cheese form might motivate to introduce the
divisor $D_3+D_8$, but as can easily be seen from
the toric data the only monomial of this degree is $x_3\cdot x_8$,
so that there is no smooth surface representing it.

\subsubsection*{Orientifold projection}

There exist two inequivalent orientifold projections $\sigma: \{x_1 \leftrightarrow -x_1, x_3 \leftrightarrow -x_3\}$
featuring that the Wilson line goldstino is in $H_{+}^1(W,\cO)$ and
$h^{11}_-({\cal M})=0$ .
However, only under the second involution $x_3 \leftrightarrow -x_3$ \
the $W$ divisor is  ${\cal O}(1)$ while for the first involution
$W$ is of $SP$-type.

\vspace{0.3cm}
\noindent
{\it Involution $\sigma: x_3 \leftrightarrow -x_3$}

\vspace{0.1cm}

The fixed point locus in this case is a bit intricate to find.
First,
there is again a component given by
$O7=D_2\sqcup D_3$. However, looking more closely
one finds that the intersection of the two divisors
$D_1\cap D_7$ gives a $\IP^2$ surface in the toric ambient
space. By the equivalence relations it is fixed under
$\sigma$. The hypersurface, in the orientifold
containing only polynomials invariant under $\sigma$, intersects
this $\IP^2$ non-generically\footnote{We are grateful to
Christoph Mayrhofer to pointing this out to us.}.
In fact, it lies already
completely on the hypersurface. Since $D_7$ intersects
the hypersurface also in a $\IP^2$, it means that these
two $\IP^2$ become identical for the $\sigma$-invariant
restricted hypersurface. Therefore, in total we
have three $O7$-components of the fixed point locus on ${\cal M}$
\eq{
             O7=D_2\sqcup D_3\sqcup D_7\, ,
}
and no $O3$-planes. Therefore, the $\IP^2$ divisors $D_3$, $D_7$  are occupied
by  components of the $O7$-plane.
Therefore,  the contribution to the $D3$-brane tadpole is
${\chi(O7)\over 4}={114+3+3\over 4}=30$.
The topological data of relevant divisors is shown in
Table \ref{tabledivsC}.

\begin{table}[ht]
  \centering
  \begin{tabular}{c|c|c}
    divisor & $(h^{00},h^{10},h^{20},h^{11})$  &  intersection curve    \\
    \hline \hline
      &    &  \\[-0.4cm]

    $D_2$ & $(1_+,0,10_+,92_+)$ & $W: C_{g=1}$ \\
    $D_3=\IP^2$ & $(1_+,0,0,1_+)$ & $W: C_{g=1}$ \\
    $D_7=\IP^2$ & $(1_+,0,0,1_+)$ &  null   \\
    \hline
    $D_8=W$ & $(1_+,1_+,0,2_+)$ & $D_2: C_{g=1} ,\ \  D_3: C_{g=1}$ \\
  \end{tabular}
  \caption{\small Divisors and their equivariant cohomology. The first three
    lines are $O7$-plane components  and the last one
    supports an $E3$ instanton.}
  \label{tabledivsC}
\end{table}

\begin{figure}[ht]
  \centering
  \includegraphics[width=0.3\textwidth]{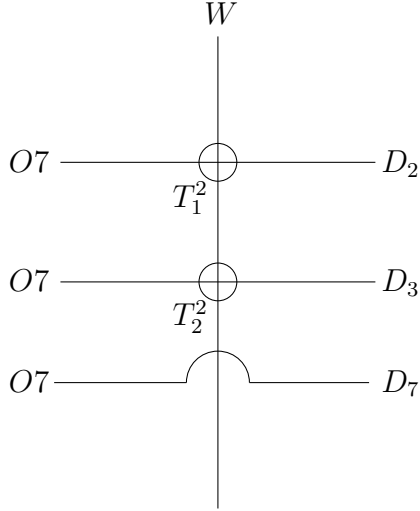}
 \begin{picture}(0,0)
   \put(-70,184){$W$}
   \put(-144,83){$O7$}
   \put(-144,45){$O7$}
   \put(-144,128){$O7$}
   \put(-3,83){$D_3$}
   \put(-3,45){$D_7$}
   \put(-3,128){$D_2$}
    \put(-82,115){$T^2_1$}
    \put(-82,70){$T^2_2$}
\end{picture}
  \caption{\small Relative position of $O7$-planes and divisor $W$ for example C with involution $\sigma: x_3 \leftrightarrow -x_3$.}
  \label{fig:oplaneB}
\end{figure}

\noindent
The relative position of $W$ to
the three components of the $O7$-plane is shown in
figure \ref{fig:oplaneB}.

\vspace{0.3cm}

\noindent
Now, we face the same situation as in Example B, namely that
the  vector-like zero modes coming from the $T^2=W\cap D_3$ intersection
cannot be made massive.
Thus, we just have the superpotential for the gaugino condensation
on the pure $SO(8)$ super Yang-Mills theories localized
on the stacks of $D7$ branes wrapping $D_3$ and $D_7$
\eq{
W= A_3 \exp(-a_3 T_3) + A_7 \exp(-a_7 T_7)\, .
}

\section{Conclusions}

In this paper we have investigated under what circumstances
there can be  poly-instanton corrections for Type IIB
orientifolds of the type $\Omega \sigma (-1)^{F_L}$ with
$O7$- and $O3$-planes. We worked out the zero mode
structure for an $E3$-instanton wrapping a surface in the
Calabi-Yau threefold. In principle, the required single additional
fermionic zero mode could arise from a deformation
or a Wilson line modulino of the surface. However, we found that
holomorphicity, i.e. the requirement the
instanton to be  $O(1)$, rules out the possibility of a deformation
zero mode. Therefore, one needs precisely one Wilson line
modulino in $H_+^1(E,{\cal O})$.

We proposed that examples of such surfaces are given
by $\IP^1$ fibrations over two-tori and presented
a couple of concrete Calabi-Yau threefolds where
such divisors appear.  Here we were concentrating on threefolds
which also contained a couple of swiss-cheese type
del Pezzo surfaces. Moreover, we
were also specifying a couple of  admissible orientifold projections
and worked out the relevant equivariant cohomology.
These models are still quite simple but they can serve
both as a proof of principle that poly-instanton corrections
are possible and as proto-type examples for further studies
on moduli stabilization and inflation. Clearly, our derivation
of the contributions to the superpotential did sensitively  depend
on the chosen configuration of tadpole canceling $D7$-branes.
As emphasized in \cite{Blumenhagen:2007sm},
the instanton zero mode structure
crucially depends on the positions of all $D7$-branes.

As an intriguing observation, we found that the volume
form took a very peculiar strong swiss-cheese like form.
The implied tree-level K\"ahler potential is of the schematic
form\footnote{Apart from the difference in the fibration part, a similar volume form has been used in \cite{Cicoli:2012cy}.}
\eq{
   K=-2\log\left( a (\tau_b)^{3\over 2}
                 - b (\tau_s)^{3\over 2}
                - c (\tau_s+\tau_w)^{3\over 2}\right)\, ,}
and the  poly-instanton generated superpotential
\eq{
     W =A\, \exp(-a_s T_s) + B\, \, \exp(-a_s \,T_s-a_w\,T_w)\, .
}
can serve as the starting point of a discussion of
moduli stabilization and inflation for this class
of models \cite{BGRS}.

\vskip1cm

\subsubsection*{Acknowledgements}
We would like to thank Christoph Mayrhofer and Timo Weigand for helpful
discussions and Michele Cicoli also for useful comments about the manuscript.
R.B. is grateful to the
Simons Center for Physics and Geometry at Stony Brook University for hospitality.
XG is supported by the MPG-CAS Joint Doctoral Promotion Programme.
PS is supported by a postdoctoral research fellowship from the Alexander von Humboldt Foundation.


\section*{Appendix}

In this appendix, we analyze one of the $K3$ fibrations
found in \cite{Cicoli:2011it} from the  new perspective
developed in this paper.

As discussed in section {\bf 2},  without
invoking additonal mechanisms,    
an instanton supported on the $K3$ fiber does not
generate  a  poly-instanton correction to the superpotential. 
However, it is still  possible to have such a correction 
as long as the underlying (fibred) Calabi-Yau
threefold contains a Wilson line  divisor being of type $O(1)$ 
with  $H_{+}^1(W,\cO)=1$ under an  appropriate holomorphic 
involution. 

As an example, consider the Calabi-Yau threefold
${\cal M}$ defined by the following  toric data
\begin{table}[ht]
  \centering
  \begin{tabular}{c|cccccccc}
     & $x_1$  & $x_2$  & $x_3$  & $x_4$  & $x_5$ & $x_6$  & $x_7$ & $x_8$        \\
    \hline
    12 & 0 & 1 & 0 & 0 & 2 & 2 &  6 & 1  \\
    6  & 0 & 0 & 0 & 1 & 1 & 1 &  3 & 0  \\
    4  & 0 & 0 & 1 & 0 & 0 & 1 &  2 & 0  \\
    6  & 1 & 1 & 0 & 0 & 1 & 0 &  3 & 0  \\
  \end{tabular}
\end{table}

\noindent
with SR-ideal:
\eq{
{\rm SR}=  \left\{ x_1\, x_2 , x_2\, x_8 , x_1\, x_3 , x_1\, x_4 , x_4\, x_5 , x_3\, x_6\, x_7 , x_5\,x_6\,x_7\,x_8
\right\}\, .
}
This Calabi-Yau has been also presented  in appendix of \cite{Cicoli:2011it},
from where we recall some of the relevant information. In the divisor basis
$\{D_1,D_2,D_3,D_4 \}$ the triple intersection form is given 
as,
\eq{ I_3= D_1^3-2D_2 D_3^2+4D_3^2+4D_3^3+2D_2D_3D_4-4D_3D_4^2. }
Expanding the K\"ahler form as $J=t_1\, D_1 + t_2\, D_2+ t_3\, D_3+ t_4\, D_4$,
the volume of the Calabi-Yau is given by
\eq{
      {\cal V}\equiv {1\over 3!}\int_{\cal M} J\wedge J\wedge J= \frac{t_1^3}{6} - t_2 t_3^2 + \frac{2 t_3^3}{3} + 2 t_2 t_3 t_4 -2 t_3 t_4^2\, .
}
$D_1$ is a shrinkable del Pezzo
${dP}_8$, $D_2$ is a  $K3$, $D_3$ is a non-shrinkable $dP_5$ and the $D_4$ is
the relevant Wilson line divisor $W$. There are four generators $\{K_i,
i=1,2,3,4\}$ for the toric K\"ahler cone and expanding the K\"ahler form  as
$J=r^i [K_i]$, the K\"ahler cone is given as $r^i>0$  for
\eq{
\label{KaehlerconeApp}
K_1 &=-3 D_1 +6 D_2 +2 D_3 +3 D_4 \, , \\
K_2 &=D_2\, ,\\
K_3 &=2 D_2 + D_4\, ,\\
K_4 &=6 D_2 + 2 D_3 +3 D_4\, .
}
The corresponding 4-cycle volumes read 
\eq{
       \tau_1&= \frac{t_1^2}{2}\, , \qquad
       \tau_2=-t_3(t_3 -2 t_4)\, , \\
       \tau_3&= 2 (t_3 -t_4)(-t_2 +t_3 +t_4)\, , \qquad
       \tau_4= 2 t_3 (t_2 -2 t_4)\, .
}
Taking into account  the K\"ahler cone constraints \eqref{KaehlerconeApp}, the volume of the Calabi-Yau can be written in terms of four-cycle volumes as
\eq{
\label{eq:volume_appendix}
{\cal V}&=- \frac{\sqrt2}{3} \tau_1^{3/2} + \frac{1}{6\sqrt2}\biggl(2(2\tau_2+2\tau_3 +\tau_4)-\sqrt{8\tau_2 \tau_3 + (2 \tau_3 +\tau_4)^2}\biggr) \\
& \hskip 4cm \times \sqrt{(2\tau_2+2\tau_3 +\tau_4)+\sqrt{8\tau_2 \tau_3 + (2
    \tau_3 +\tau_4)^2}} \; .
}
The above volume form is quite complicated and one way to define the large
volume limit is taking the large fiber limit $\tau_2\rightarrow \infty$ while
keeping the  other divisor volumes to be relatively small. In this limit, 
the above volume expression (\ref{eq:volume_appendix}) reduces to 
\eq{
{\cal V}|_{\tau_2\rightarrow \infty}={\textstyle \frac{2}{3}} \tau_2^{3/2}+{\textstyle \frac{1}{2}}
(\tau_3+\tau_4){\sqrt{\tau_2}} - {\textstyle \frac{\sqrt2}{3}} \tau_1^{3/2} +
{\textstyle \frac{1}{3\sqrt 2}} \tau_3^{3/2} +
{\cal O}({\tau_2}^{-1/2}).
}
We observe that the typical volume factor for a $K3$ fibration
$(\tau_3+\tau_4){\sqrt{\tau_2}}$ indeed appears and that
it now also contains the volume of the Wilson line divisor $\tau_4$. 

There exist two inequivalent orientifold projections $\sigma: \{x_2 \leftrightarrow -x_2, x_5 \leftrightarrow -x_5\}$
featuring that the Wilson line Goldstino is in $H_{+}^1(W,\cO)$ and
$h^{11}_-({\cal M})=0$. It can also be checked that the $D3$ and $D7$-branes
tadpoles can be canceled. The topological data of the relevant divisors are
shown in  Table \ref{tabledivsaa} and Table \ref{tabledivsab}. 
Finally, it can be shown that for both
involutions there are no vector-like zero modes on the intersection 
 $E3\cap D7$ and one gets the following form of the superpotential,
\eq{
 W =A_1\, \exp(-2\pi T_1) + A_1 A_4 \, \exp(-2\pi \,T_1-2\pi\,T_4)\, .
}

\begin{table}[ht]
  \centering
  \begin{tabular}{c|c|c}
    divisor &$(h^{00},h^{10},h^{20},h^{11})$ & intersection curve    \\
    \hline
      &     \\[-0.4cm]
    $D_2=K3$  &$(1_+,0,1_+,20_+)$ & $W: C_{g=1}$ \\
    $D_8=dP_{10}$  & $(1_+,0,0,11_+)$ & $W: C_{g=1}$,\ \ $D_1: C_{g=1}$\\
  \hline
    $D_1=dP_8$ & $(1_+,0,0,5_++4_-)$ & $D_8: C_{g=1}$   \\
    $D_4=W$ &$(1_+,1_+,0,2_+)$ &  $D_2: C_{g=1} , \ \ D_8: C_{g=1}$ \\
  \end{tabular}
  \caption{\small Divisors and their equivariant cohomology. Under $x_2 \leftrightarrow -x_2$. The first two
    lines are $O7$-plane components  and the $D_1$, $D_4$ divisors
    support  $E3$ instanton.}
  \label{tabledivsaa}
\end{table}

\begin{table}[ht]
  \centering
  \begin{tabular}{c|c|c}
    divisor &$(h^{00},h^{10},h^{20},h^{11})$& intersection curve    \\
    \hline
      &     \\[-0.4cm]
    $D_5$  &$(1_+,0,2_+,31_+)$ &  $D_1: C_{g=1}$ \\
  \hline
    $D_1=dP_8$ & $(1_+,0,0,5_++4_-)$ & $D_5: C_{g=1}$   \\
    $D_4=W$ & $(1_+,1_+,0,2_+)$&  $D_2: C_{g=1} $ \\
  \hline
    $D_2=K3$  & $(1_+,0,1_-,10_++10_-)$ & $W: C_{g=1}$ \\
  \end{tabular}
  \caption{\small Divisors and their equivariant cohomology. Under $x_5 \leftrightarrow -x_5$. The first line is an $O7$-plane component  and the $D_1$, $D_4$ divisors
    support $E3$ instanton.}
  \label{tabledivsab}
\end{table}
\noindent


\clearpage
\bibliography{rev2}
\bibliographystyle{utphys}


\end{document}